\font\bbb=msbm10                                                   

\def\C{\hbox{\bbb C}}

\def\R{\hbox{\bbb R}}
\def\Z{\hbox{\bbb Z}}

\def\APS{{\sl Acta Physica Slovaca}}

\def\CPAM{{\sl Commun.\ Pure Appl.\ Math.}}

\def\CS{{\sl Complex Systems}}

\def\FP{{\sl Found.\ Phys.}}

\def\IJTP{{\sl Int.\ J. Theor.\ Phys.}}

\def\JSP{{\sl J. Stat.\ Phys.}}

\def\PD{{\sl Physica D}}

\def\PLA{{\sl Phys.\ Lett.\ A}}

\def\PRA{{\sl Phys.\ Rev.\ A}}

\def\PRD{{\sl Phys.\ Rev.\ D}}
\def\PRE{{\sl Phys.\ Rev.\ E}}

\def\RMP{{\sl Rev.\ Mod.\ Phys.}}
\def\Sc{{\sl Science}}
\def\SPJETP{{\sl Sov.\ Phys.\ JETP}}

\def\dajm{\hbox{D. A. Meyer}}

\def\brosl{\hbox{B. Hasslacher}}
\def\bd{\hbox{\brosl\ and \dajm}}

\def\feynman{\hbox{R. P. Feynman}}

\def\gz{\hbox{G. Gr\"ossing and A. Zeilinger}}

\def\hfb{\hfil\break}

\catcode`@=11
\newskip\ttglue

   \font\ninerm=cmr9    \font\eightrm=cmr8   \font\sixrm=cmr6
  \font\ninebf=cmbx9   \font\eightbf=cmbx8  \font\sixbf=cmbx6
  \font\nineit=cmti9   \font\eightit=cmti8  
  \font\ninesl=cmsl9   \font\eightsl=cmsl8  
  \font\ninemi=cmmi9   \font\eightmi=cmmi8  \font\sixmi=cmmi6

\font\bigtenbf=cmr10 scaled\magstep2 

\def\ninepoint{\def\rm{\fam0\ninerm}%
  \textfont0=\ninerm \scriptfont0=\sixrm
  \textfont1=\ninemi \scriptfont1=\sixmi
  \textfont\itfam=\nineit  \def\it{\fam\itfam\nineit}%
  \textfont\slfam=\ninesl  \def\sl{\fam\slfam\ninesl}%
  \textfont\bffam=\ninebf  \scriptfont\bffam=\sixbf
    \def\bf{\fam\bffam\ninebf}%
  \tt \ttglue=.5em plus.25em minus.15em
  \normalbaselineskip=11pt
  \setbox\strutbox=\hbox{\vrule height8pt depth3pt width0pt}%
  \normalbaselines\rm}

\def\eightpoint{\def\rm{\fam0\eightrm}%
  \textfont0=\eightrm \scriptfont0=\sixrm
  \textfont1=\eightmi \scriptfont1=\sixmi
  \textfont\itfam=\eightit  \def\it{\fam\itfam\eightit}%
  \textfont\slfam=\eightsl  \def\sl{\fam\slfam\eightsl}%
  \textfont\bffam=\eightbf  \scriptfont\bffam=\sixbf
    \def\bf{\fam\bffam\eightbf}%
  \tt \ttglue=.5em plus.25em minus.15em
  \normalbaselineskip=9pt
  \setbox\strutbox=\hbox{\vrule height7pt depth2pt width0pt}%
  \normalbaselines\rm}

\def\sfootnote#1{\edef\@sf{\spacefactor\the\spacefactor}#1\@sf
      \insert\footins\bgroup\eightpoint
      \interlinepenalty100 \let\par=\endgraf
        \leftskip=0pt \rightskip=0pt
        \splittopskip=10pt plus 1pt minus 1pt \floatingpenalty=20000
        \parskip=0pt\smallskip\item{#1}\bgroup\strut\aftergroup\@foot\let\next}
\skip\footins=12pt plus 2pt minus 2pt
\dimen\footins=30pc

\def\ie{{\it i.e.}}
\def\eg{{\it e.g.}}

\def\etal{{\it et al.}}

\magnification=1200
\input epsf.tex

\dimen0=\hsize \divide\dimen0 by 13 \dimendef\chasm=0
\dimen1=\hsize \advance\dimen1 by -\chasm \dimendef\usewidth=1
\dimen2=\usewidth \divide\dimen2 by 2 \dimendef\halfwidth=2
\dimen3=\usewidth \divide\dimen3 by 3 \dimendef\thirdwidth=3
\dimen4=\hsize \advance\dimen4 by -\halfwidth \dimendef\secondstart=4
\dimen5=\halfwidth \advance\dimen5 by -10pt \dimendef\indenthalfwidth=5
\dimen6=\thirdwidth \multiply\dimen6 by 2 \dimendef\twothirdswidth=6
\dimen7=\twothirdswidth \divide\dimen7 by 4 \dimendef\qttw=7
\dimen8=\qttw \divide\dimen8 by 4 \dimendef\qqttw=8
\dimen9=\qqttw \divide\dimen9 by 4 \dimendef\qqqttw=9

\parskip=0pt
\line{\hfil July 1996}
\line{\hfil{\it revised\/} October 1996}
\line{\hfil quant-ph/9611005}
\bigskip\bigskip\bigskip
\centerline{\bf\bigtenbf QUANTUM MECHANICS OF LATTICE GAS AUTOMATA}
\bigskip
\centerline{\bf\bigtenbf I.  ONE PARTICLE PLANE WAVES}
\bigskip
\centerline{\bf\bigtenbf AND POTENTIALS}
\vfill
\centerline{\bf David A. Meyer}
\bigskip
\centerline{\sl Institute for Physical Sciences}
\smallskip
\centerline{\sl and}
\smallskip
\centerline{\sl Project in Geometry and Physics}
\centerline{\sl Department of Mathematics}
\centerline{\sl University of California/San Diego}
\centerline{\sl La Jolla, CA 92093-0112}
\centerline{dmeyer@chonji.ucsd.edu}
\vfill
\centerline{ABSTRACT}
\bigskip
\noindent Classical lattice gas automata effectively simulate physical 
processes such as diffusion and fluid flow (in certain parameter
regimes) despite their simplicity at the microscale.  Motivated by 
current interest in quantum computation we recently defined 
{\sl quantum\/} lattice gas automata; in this paper we initiate a 
project to analyze which physical processes these models can 
effectively simulate.  Studying the single particle sector of a one
dimensional quantum lattice gas we find discrete analogues of plane
waves and wave packets, and then investigate their behaviour in the
presence of inhomogeneous potentials.

\bigskip
\global\setbox1=\hbox{PACS numbers:\enspace}
\global\setbox2=\hbox{PACS numbers:}
\parindent=\wd1
\item{PACS numbers:}  03.65.-w,  
                      02.70.-c,  
                      11.55.Fv,  
                      89.80.+h.  
\item{\hbox to \wd2{KEY\hfill WORDS:}}   
                      quantum lattice gas; quantum cellular automaton; 
                      quantum computation.

\vfill
\eject

\headline{\ninepoint\it Quantum mechanics of LGA I.
          \hfil David A. Meyer}
\parskip=10pt
\parindent=20pt

\noindent{\bf 1.  Introduction}

\noindent  The first quantum lattice gas automaton (QLGA) appeared as
Feynman's path integral for a relativistic particle in $1+1$ 
dimensions [1]; independently Riazanov constructed a $2+1$ dimensional
QLGA as the path integral for the next higher dimensional Dirac
equation [2].  In these formulations the quantum particle is 
conceptualized as evolving along spacetime trajectories, each of which
is assigned a probability amplitude which is the product of a sequence
of `scattering' amplitudes describing the evolution of the particle 
during a single timestep.  Thus these QLGA are discretizations of 
quantum mechanical processes.

Feynman's path integral formulation of quantum mechanics reproduces
the standard Schr\"odinger formulation of wave functions obeying
partial differential equations [1].  These differential equations can
be discretized directly, giving equations which Succi and Benzi 
naturally identify in the lattice gas paradigm as quantum lattice
Boltzmann equations [3].  It is a familiar, although not often 
useful, observation that any numerical evolution of a discretized 
partial differential equation can be interpreted as the evolution of 
some cellular automaton (CA), if one allows the set of states to be
$\R$, or $\C$, or $\Z_N$ for some very large $N$.  Taking this 
perspective, Bialynicki-Birula constructs a model for quantum 
evolution---a linear unitary CA [4]---which is essentially equivalent 
to, although derived independently of, Succi and Benzi's equations.

The equivalence of a QLGA simulation and the evolution of a set of 
quantum lattice Boltzmann equations/unitary CA depends on the 
equivalence of the path integral and standard formulations of quantum 
mechanics in the continuum.  Our recent work explaining the necessity 
of non-unitarity in earlier attempts of Gr\"ossing, Zeilinger, \etal\ 
to construct homogeneous CA for quantum evolution [5] demonstrates 
this equivalence directly for the discrete models [6].  We also note 
that, in contrast to simulation with deterministic or probabilistic 
LGA, simulation with a QLGA {\sl requires\/} evolution along all 
possible spacetime trajectories.  This may be achieved (slowly) by
evolution of the quantum lattice Boltzmann equation on a classical
computer or, at present hypothetically (but rapidly), by simulation
on a quantum computer.

In fact, given the arguments that massive parallelism will optimize 
nanoscale quantum computer architecture [7], it is plausible that the
first useful {\sl quantum computation\/} [8] will implement a QLGA 
simulation of some quantum mechanical process.  This provides two 
reasons to pursue the project described in this series of papers:  we
want to explore not only quantum mechanical phenomena which can be 
simulated effectively by QLGA, but also how well, as Feynman 
suggested [9], a quantum computer might simulate physics.  In 
addition, we expect the quantum mechanics of LGA to have implications
for discrete models of fundamental physics:  we have already found
remarkable consequences of unitarity in linear [6,10] and nonlinear
[11] QCA.

We begin in the next section by recalling the model of [6] with which 
we will be working:  the most general one dimensional homogeneous QLGA
with a single particle of speed no more than 1 in lattice units.  The 
local evolution rule for this model has {\sl two\/} free parameters:
essentially the second measures the coupling between two copies of
Feynman's original QLGA in which the first measures the `mass' of the
particle.

This generalized QLGA is exactly solvable, just as is a single Feynman
QLGA.  In Section 3 we demonstrate this by finding the discrete
analogues of plane waves in, and the dispersion relation for, our
QLGA.  We also show the results of simulations of the former---on a
{\sl deterministic\/} computer.

We might imagine a one particle QLGA being simulated {\sl quantum 
mechanically\/} by a ballistic electron in a solid state lattice [12] 
or as the `low energy' sector of a line of dynamical quantum spins 
[9] (`low energy' meaning, \eg, the configurations with one spin up 
and the rest down).  In the former case [13], and certainly if our 
interest is in the QLGA as a discrete approximation to the Dirac 
equation [6], it is natural to investigate wave packets representing
a semi-classical quantum particle.  We do so in Section 4.

In Section 5 we show how to introduce an inhomogeneous potential into
the model.  Concentrating on finite square well potentials, we 
determine the dependence of the frequency/energy eigenvalues on the
depth of the well and find that the eigenfunctions take the expected
form.  Finally, we utilize the results of Section 4 and show the 
results of simulations of a wave packet in a finite square well.

We summarize our results in Section 6 and indicate the directions in
which this research is continuing.

\medskip
\noindent{\bf 2.  The one particle QLGA}
\nobreak

\nobreak
\noindent A lattice gas automaton (LGA) should be envisioned as a 
collection of particles moving synchronously from vertex to vertex on 
a fixed graph (lattice) $L$:  At the beginning of each timestep each 
particle is located at some vertex and is labelled with a `velocity' 
indicating along which edge incident to that vertex it will move 
during the `advection' half of the timestep.  After moving along the 
designated edge to the next vertex, in the `scattering' half of the 
timestep the particles at each vertex interact according to some rule 
which assigns new `velocity' labels to each.  For the purposes of this 
paper we will consider only one dimensional lattices $L$, isomorphic 
to the integer lattice $\Z$ or some periodic quotient thereof.  In 
this case there are only two possible `velocities':  left and right.  
We will further restrict our attention to LGA with only a single 
particle; for some preliminary work on QLGA with multiple particles, 
see [6,14,15].

A QLGA is a LGA for which the time evolution is unitary.  To make this
precise we must first identify the Hilbert space of the theory.  For a
one particle QLGA in one dimension an orthonormal basis for the 
Hilbert space $H$ is given by $|x,\alpha\rangle$ (in the standard 
Dirac notation [16]), where $x \in L$ denotes position and 
$\alpha \in \{\pm 1\}$ denotes `velocity'.  At each time the state of 
the QLGA is described by a {\sl state vector\/} in $H$:
$$
\Psi(t) = \sum_{x,\alpha} \psi_{\alpha}(t,x) |x,\alpha\rangle,
                                                            \eqno(2.1)
$$
where the {\sl amplitudes\/} $\psi_{\alpha}(t,x) \in \C$ and the norm 
of $\Psi(t)$, as measured by the inner product on $H$, is:  
$$
1 = \sum_{x,\alpha} \overline{\psi_{\alpha}(t,x)} \psi_{\alpha}(t,x).
                                                            \eqno(2.2)
$$
The state vector evolves unitarily, \ie, $\Psi(t+1) = U \Psi(t)$, 
where $U$ is a unitary operator on $H$.  Since the evolution is 
unitary, the inner product is preserved and (2.2) holds for all times 
if it holds for one; this allows the interpretation of 
$\overline{\psi_{\alpha}(t,x)} \psi_{\alpha}(t,x)$ as the 
{\sl probability\/} that the particle be in the state 
$|x,\alpha\rangle$ at time $t$ [16,17].  As usual, therefore, the 
basis state vectors $\Psi = |x,\alpha\rangle$ correspond to 
`classical' states---with probability 1 there is a single particle at
$x$ with `velocity' $\alpha$---and a generic state vector (2.1) is a 
superposition of these `classical' states, each of which has integer
values (one 1, the rest 0) for the number of particles at each lattice
site.  (In general, the basis vectors for the $n$ particle subspace of 
the QLGA Hilbert space are exactly the possible states of a classical
deterministic LGA with $n$ particles [15].)

In order for the evolution to have the `advection' interpretation
described above, the basis vectors should evolve so that
$$
\langle x,\alpha|U|y,\beta \rangle \not= 0                  \eqno(2.3)
$$
{\sl only\/} when $x = y + \beta$.  This is equivalent to a condition 
on the amplitudes:
$$
\psi_{\alpha}(t+1,x) 
 = f_{\alpha}\bigl(\psi_{-1}(t,x+1),\psi_{+1}(t,x-1)\bigr),
$$
where taking $f_{\alpha}$ to be independent of $x$ means that the QLGA
is {\sl homogeneous\/} in space.  As the notation suggests, it is 
convenient to combine the left and right moving amplitudes at $x$ into 
a two component complex vector 
$\psi(t,x) := \bigl(\psi_{-1}(t,x),\psi_{+1}(t,x)\bigr)$ so that a 
state vector is written
$$
\Psi(t) = \sum_x \psi(t,x) |x\rangle.
$$
We showed in [6] that the most general unitary evolution for a one 
dimensional QLGA with parity invariance%
\sfootnote*{{\it I.e.}, invariance under $x \to -x$; also called
            {\sl reflection\/} invariance.}
is unitarily equivalent to
$$
\psi(t+1,x)
=
\pmatrix{0 & i\sin\theta \cr
         0 &  \cos\theta \cr
        }
\psi(t,x-1)
+
\pmatrix{ \cos\theta & 0 \cr
         i\sin\theta & 0 \cr
        }
\psi(t,x+1),                                                \eqno(2.4)
$$
up to some overall phase which has no physical effect.  Here the 
parameter $\theta \in \R$, or more precisely, $\tan\theta$, plays a 
role something like `mass':  when $\theta = 0$ the particle travels 
only on the lightcone; as $\tan\theta$ increases its probability for 
moving more slowly does also.

Notice that just as in deterministic LGA in one dimension, the 
particle has a $\Z_2$ valued `Lagrangian' conserved quantity measuring
the parity of its fiducial space coordinate.  This `spurious' 
conserved quantity partitions the set of particles in a deterministic
LGA into two decoupled gases [18] and in the QLGA defined by (2.4) it
partitions the set of amplitudes $\psi(t,x)$ into two independent sets
according to $x+t$ (mod 2).  This motivates consideration of the most
general, no less local model which breaks this symmetry, namely
$$
\psi(t+1,x) 
= w_{-1}\psi(t,x-1) + w_0\psi(t,x) + w_{+1}\psi(t,x+1),     \eqno(2.5)
$$
where $w_i \in M_2(\C)$ are $2\times 2$ complex matrices.  In terms of
the basis vectors $|x,\alpha\rangle$, now (2.3) holds when 
$x = y + \beta$ {\sl or\/} $x = y$, \ie, the particle can have nonzero 
amplitude to maintain its position.  The condition that the global 
evolution, \ie, the matrix $U$, be unitary is expressed in terms of 
the $w_i$ by the equations:
$$
\eqalignno{
w_{-1}^{\vphantom{\dagger}}w_{-1}^{\dagger} + 
w_0^{\vphantom{\dagger}}w_{0\phantom-}^{\dagger} + 
w_{+1}^{\vphantom{\dagger}}w_{+1}^{\dagger}         &= I           \cr
w_0^{\vphantom{\dagger}}w_{-1}^{\dagger} + 
w_{+1}^{\vphantom{\dagger}}w_{0\phantom+}^{\dagger} &= 0
                                                             &(2.6)\cr
w_{+1}^{\vphantom{\dagger}}w_{-1}^{\dagger}         &= 0,          \cr
}
$$
together with their Hermitian conjugates [6].  Imposing also the 
condition of parity (reflection) invariance on evolution of the form 
(2.5), we showed in [6] that the most general solution, up to unitary 
equivalence and an overall phase, is given by
$$
\setbox1=\hbox{%
$
w_{-1} = \cos\rho\pmatrix{0 & i\sin\theta \cr
                          0 &  \cos\theta \cr
                         }
\qquad
w_{+1} = \cos\rho\pmatrix{ \cos\theta & 0 \cr
                          i\sin\theta & 0 \cr
                         }
$
}
\eqalign{
w_{-1} = \cos\rho\pmatrix{0 & i\sin\theta \cr
                          0 &  \cos\theta \cr
                         }
\qquad
w_{+1} = \cos\rho\pmatrix{ \cos\theta & 0 \cr
                          i\sin\theta & 0 \cr
                         }                                         \cr
\hbox to\wd1{\hfil%
$
w_0  =   \sin\rho\pmatrix{  \sin\theta & -i\cos\theta \cr
                          -i\cos\theta &   \sin\theta \cr
                         }.
$ 
\hfil}                                                             \cr
}                                                           \eqno(2.7)
$$
Here $\rho \in \R$ is a coupling parameter breaking the `spurious' 
symmetry.  When $\rho = 0$, (2.5) reduces to (2.4), the QLGA which is 
unitarily equivalent to the models of Feynman [1], Succi and Benzi 
[3], and Bialynicki-Birula [4].  As $\tan\rho$ increases, the relative 
weight of $w_0$ increases and the particle has greater probability of 
maintaining its position, \ie, having zero velocity.  This is the 
first indication of a symmetry between $\theta$ and $\rho$ which will 
become more explicit as we investigate the general QLGA of (2.5) and 
(2.7).

\medskip
\noindent{\bf 3.  Plane waves}
\nobreak

\nobreak
\noindent The local evolution rule (2.5) is linear so we expect the
model to be exactly solvable.  In [6] we solved the $\rho = 0$ case 
by counting spacetime lattice paths in order to compute the 
{\sl propagator\/} 
$K_{\alpha\beta}(t,x;0,0) := \langle x,\alpha|U^t|0,\beta \rangle$
explicitly.  Lattice paths are more difficult to count when the 
particle has nonzero amplitudes for maintaining its position during
each timestep.  Avoiding this difficulty leads us to a more physical 
approach---finding the discrete analogue of plane waves in a QLGA.

Recall that the QLGA is homogeneous, \ie, $U$ commutes with the 
translation (shift) operator $T$ defined by $(T\psi)(x) := \psi(x+1)$.  
Suppose $L = \Z_N$.  Then the eigenvalues of $T$ are $e^{ik}$ for 
{\sl wave numbers\/} $k = 2\pi n/N$, $n \in \{0,\ldots,N-1\}$, and the 
corresponding eigenvectors $\Psi^{(k)}$ satisfy:
$$
\psi^{(k)}(x+1) = e^{ik} \psi^{(k)}(x).                     \eqno(3.1)
$$
Since $[U,T] = 0$ and $U$ is unitary, the $\Psi^{(k)}$ are also 
eigenvectors for $U$ with
$$
U \Psi^{(k)} = e^{-i\omega_k} \Psi^{(k)},                   \eqno(3.2)
$$
for some {\sl frequencies\/} $\omega_k \in \R$.  The eigenvectors 
$\Psi^{(k)}$ are the discrete analogues of plane waves since they 
evolve simply by phase multiplication.

Since the action of $U$ is defined by (2.5), (3.1) and (3.2) imply 
that 
$$
\eqalign{
e^{-i\omega_k} \psi^{(k)}(x) 
 &= w_{-1} \psi^{(k)}(x-1) 
  + w_0 \psi^{(k)}(x)
  + w_{+1} \psi^{(k)}(x+1)                                         \cr
 &= (e^{-ik} w_{-1} + w_0 + e^{ik} w_{+1}) \psi^{(k)}(x)           \cr
 &=: D(k) \psi^{(k)}(x).                                           \cr
}                                                           \eqno(3.3)
$$
Thus the $e^{-i\omega_k}$ are eigenvalues of $D(k) \in M_2(\C)$, \ie,
solutions of
$$
\det (D(k) - e^{-i\omega_k} I) = 0,                         \eqno(3.4)
$$
where $I$ is the $2\times 2$ identity matrix.  Using the 
parametrization (2.7) of the $w_i$, (3.4) reduces to the condition
$$
\cos\omega = \cos k \cos\theta \cos\rho + \sin\theta \sin\rho.
                                                            \eqno(3.5)
$$
For a given wave number $k$, (3.5) determines two frequencies 
$\pm\omega_k$ in terms of the rule parameters $\theta$ and $\rho$.  
Call the corresponding eigenvectors of $D(k)$ (normalized to have 
length $1/N$) $\psi^{(k,\pm1)}(0) \in \C^2$, so that the corresponding 
plane waves are defined by (3.1) to be
$$
\Psi^{(k,\epsilon)} 
 := \sum_x \psi^{(k,\epsilon)}(0) e^{ikx} |x\rangle.        \eqno(3.6)
$$
Figures 1 and 2 show the evolution of $\epsilon = +1$ (right moving) 
plane waves for $n = 1,2$.  The probability 
$\psi^{\dagger}(t,x) \psi(t,x)$ (where 
$\psi^{\dagger}(t,x) := {}^t\overline{\psi(t,x)}$) of the particle 
being at $x$ is constant in $x$ (and $t$), so the vertical axis in the
graphs shows the real part of $\psi_{-1}(t,x)$.  Even on such a small 
($N = 32$) lattice this QLGA provides a very good approximation to 
continuum plane waves of long wavelength measured in lattice units.

\topinsert
\null\vskip-\baselineskip
\vskip-\baselineskip
$$
\epsfxsize=\halfwidth\epsfbox{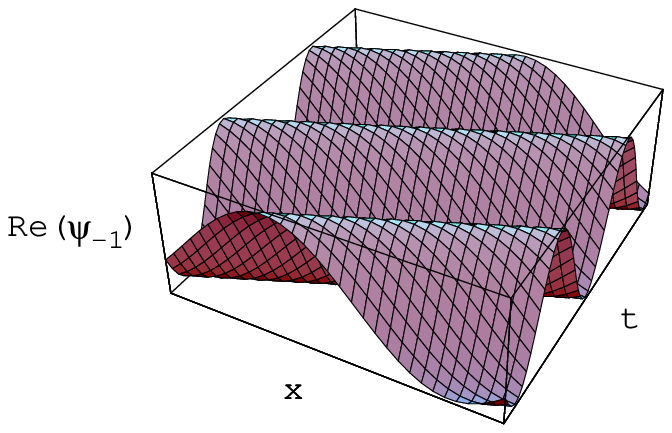}\hskip\chasm%
\epsfxsize=\halfwidth\epsfbox{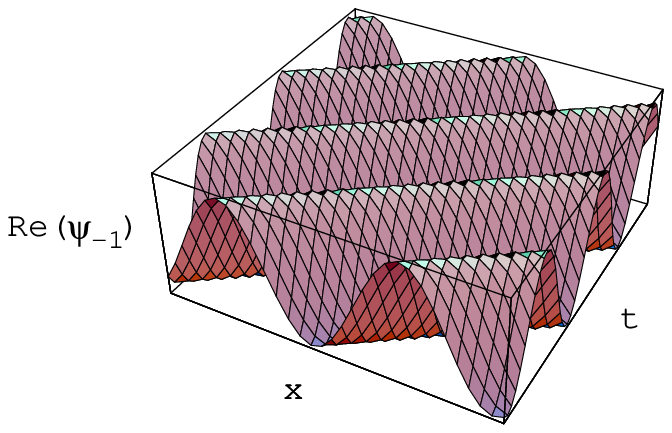}
$$
\hbox to\hsize{%
\vbox{\hsize=\halfwidth\eightpoint{%
\noindent{\bf Figure 1}.  Evolution of the $n = 1$ right moving plane 
wave on a periodic lattice with $\theta = \pi/3$ and $\rho = \pi/4$.
}}
\hfill%
\vbox{\hsize=\halfwidth\eightpoint{%
\noindent{\bf Figure 2}.  Evolution of the $n = 2$ right moving plane 
wave on a periodic lattice with $\theta = \pi/3$ and $\rho = \pi/4$.
}}}
\endinsert

Notice that when the wave number $k$ increases, so does the frequency 
$\omega$---the time period is shorter in Figure 2 than in Figure 1.  
Also the {\sl phase velocity\/} $\omega/k$ decreases---the crest of 
the wave moves more slowly.  In fact, (3.5) is the exact 
{\sl dispersion relation}, giving the frequency in terms of the wave 
number.  Figure 3 graphs the dispersion relation for the QLGA with the 
same rule parameters used in the simulations of Figures 1 and 2.  The 
graph has reflection symmetry about both axes since (3.5) is invariant 
under both $k \to -k$ and $\omega \to -\omega$.  Each reflection alone 
changes the direction of the plane wave.  When $k = 0$, 
$\omega = \pm(\theta - \rho)$; when $k = \pm\pi$, 
$\omega = \pm(\theta + \rho - \pi)$.  These values exemplify another 
symmetry of the dispersion relation---invariance under 
$\theta \longleftrightarrow \rho$; this is a symmetry in the QLGA rule 
space which is not realizable by a {\sl local\/} unitary 
transformation.  Figure 4 graphs the special case of equal rule
parameters; here the dispersion relation passes through the origin.

\topinsert
\null\vskip-\baselineskip
\vskip-\baselineskip
$$
\epsfxsize=\halfwidth\epsfbox{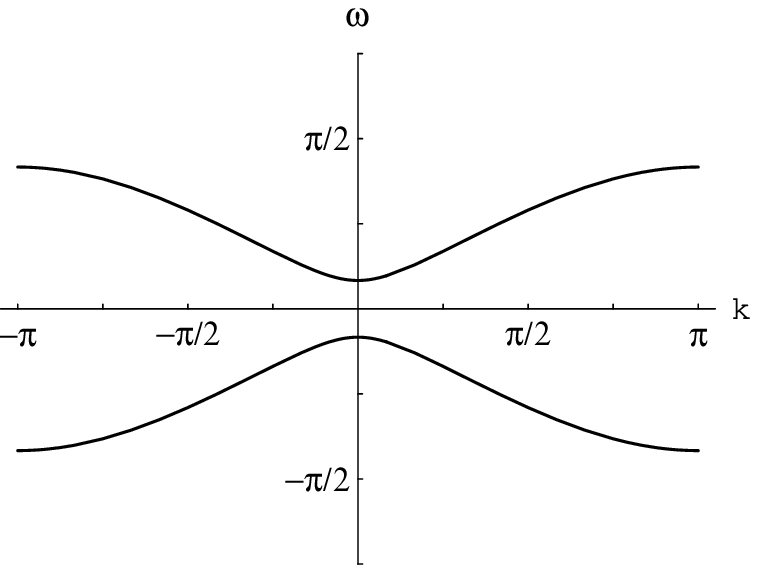}\hskip\chasm%
\epsfxsize=\halfwidth\epsfbox{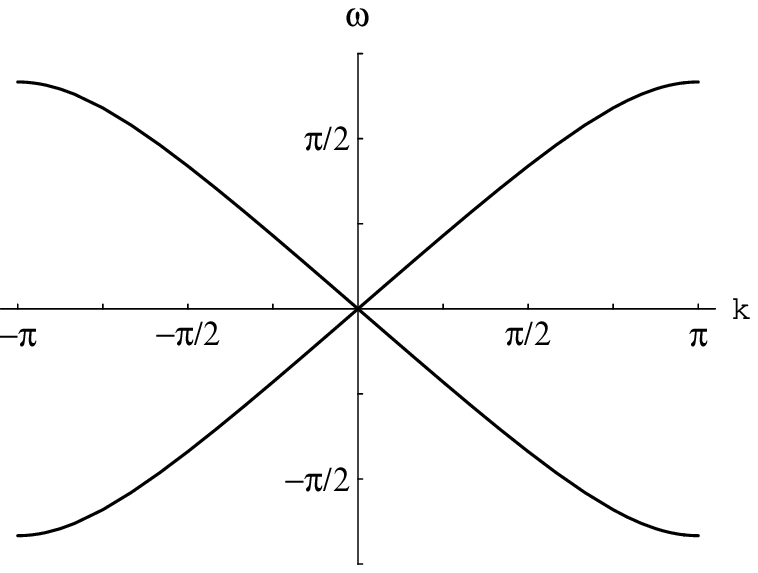}
$$
\hbox to\hsize{%
\vbox{\hsize=\halfwidth\eightpoint{%
\noindent{\bf Figure 3}.  The dispersion relation for $\theta = \pi/3$
and $\rho = \pi/4$.  $\pi/12 \le |\omega| \le 5\pi/12$.
}}
\hfill%
\vbox{\hsize=\halfwidth\eightpoint{%
\noindent{\bf Figure 4}.  The dispersion relation in the `massless' 
case $\theta = \rho = \pi/6$.  $|\omega| \le 2\pi/3$.
}}}
\endinsert

By comparison with plane waves in continuum quantum mechanics [16,17], 
we know that $\omega$ and $k$ should be interpreted as being 
proportional to {\sl energy\/} and {\sl momentum}, respectively.
Expanding the dispersion relation (3.5) around $k = 0$ and 
$\omega = 0$ to second order, we find
$$
\omega^2 
 = k^2\cos\theta\cos\rho + 2\bigl(1 - \cos(\theta - \rho)\bigr).
                                                            \eqno(3.7)
$$
For a relativistic particle in the continuum,
$$
E^2 = p^2 c^2 + m^2 c^4.                                    \eqno(3.8)
$$
Comparing (3.7) and (3.8) suggests that the 
$1 - \cos(\theta - \rho) = 0$ case, \ie, the $\theta = \rho$ case 
shown in Figure 4, corresponds to the particle being {\sl massless}.

Not only do the plane wave parameters $\omega$ and $k$ bear the 
interpretation of proportionality to the conserved quantities energy
and momentum, but they also label a complete set of (nonlocal) 
conserved quantities for the QLGA.  Since $T$ is orthogonal its
eigenvectors $\Psi^{(k,\epsilon)}$ are orthogonal for distinct wave
numbers $k$.  Furthermore, $D(k)$ is unitary, so its eigenvectors
$\psi^{(k,\pm1)}(0)$ are orthogonal for {\sl each\/} $k$ and hence so 
are the plane waves $\Psi^{(k,\pm1)}$.  Since we normalized the 
eigenvectors of $D(k)$ to have length $1/N$, the plane waves (3.6) 
form an orthonormal basis for $H$ which we denote by 
$\{|k,\epsilon\rangle\}$.  Consider any state vector $\Psi \in H$:
$$
\eqalign{
\Psi 
 &= \sum_x \psi(x) |x\rangle                                       \cr
 &= \sum_{\vphantom{k}x} \psi(x) 
    \sum_{k,\epsilon} |k,\epsilon\rangle \langle k,\epsilon|x\rangle
                                                                   \cr
 &= \sum_{k,\epsilon} 
    \Bigl(\sum_{\vphantom{k}x} 
          \langle k,\epsilon|x\rangle \psi(x)
    \Bigr) 
    |k,\epsilon\rangle                                             \cr
}
$$
The parenthesized expression is the amplitude of $|k,\epsilon\rangle$
in the new basis:
$$
\eqalignno{
\hat\psi_{\epsilon}(k) 
 &:= \sum_x \langle k,\epsilon|x\rangle \psi(x)                    \cr
 &\phantom{:}= \sum_x \Bigl(\sum_y \psi^{(k,\epsilon)}(0) e^{iky}
                                   |y\rangle
                      \Bigr)^{\dagger} |x\rangle \psi(x)           \cr
 &\phantom{:}= \bigl(\psi^{(k,\epsilon)}(0)\bigr)^{\dagger} 
               \sum_x \psi(x) e^{-ikx}                       &(3.9)\cr
 &\phantom{:}=: \bigl(\psi^{(k,\epsilon)}(0)\bigr)^{\dagger} 
                \hat\psi(k),                                       \cr
}
$$
where $\hat\psi(k)$ is the {\sl discrete Fourier transform\/} of 
$\psi(x)$.  The plane waves $|k,\epsilon\rangle$ evolve by phase 
multiplication so the probabilities 
$\overline{\hat\psi_{\epsilon}(k)}\hat\psi_{\epsilon}(k)$ are left
invariant by the evolution.  Since any initial state vector $\Psi(0)$
can be expressed in the plane wave basis this way, the existence of
these conserved quantities is equivalent to exact solvability for 
this model of a one particle QLGA.

\medskip
\noindent{\bf 4.  Wave packets}
\nobreak

\nobreak
\noindent The plane waves (3.6) provide a starting point for 
constructing wave packets with localized position and particularized 
momentum.  Consider the right moving plane wave with wave number $k_0$ 
in the position basis:
$$
\Psi^{(k_0,+1)} = \sum_x \psi^{(k_0,+1)}(0) e^{ik_0 x}|x\rangle.
$$
In this discrete (and periodic) situation the binomial distribution is 
a convenient substitute for a Gaussian distribution, so to localize 
the particle we multiply the amplitudes by appropriate binomial 
coefficients:  Let
$$
\Psi 
 := {2s \choose s}^{\!\! -1}\sum_x \psi^{(k_0,+1)}(0) e^{ik_0 x}
                              {s \choose x - x_0 + s/2}|x\rangle,
                                                            \eqno(4.1)
$$
for even $s \le N$, where the inverse binomial coefficient outside 
the sum is the requisite normalization factor.  This wave packet is 
localized around $x_0$, having support on the interval 
$[x_0 - s/2,x_0 + s/2]$.  Figure 5 shows the evolution for wave 
number $k_0 = \pi/4$ and width $s = 32$ on the lattice $\Z_{64}$.  
The rule parameters are the same as those used in the simulations 
shown in Figures 1 and 2.  In contrast to those graphs, the vertical 
axis in Figure 5 shows the probability that the particle is in the 
state $|x\rangle$.

\topinsert
\null\vskip-\baselineskip
\vskip-\baselineskip
\vskip-\baselineskip
$$
\epsfxsize=\usewidth\epsfbox{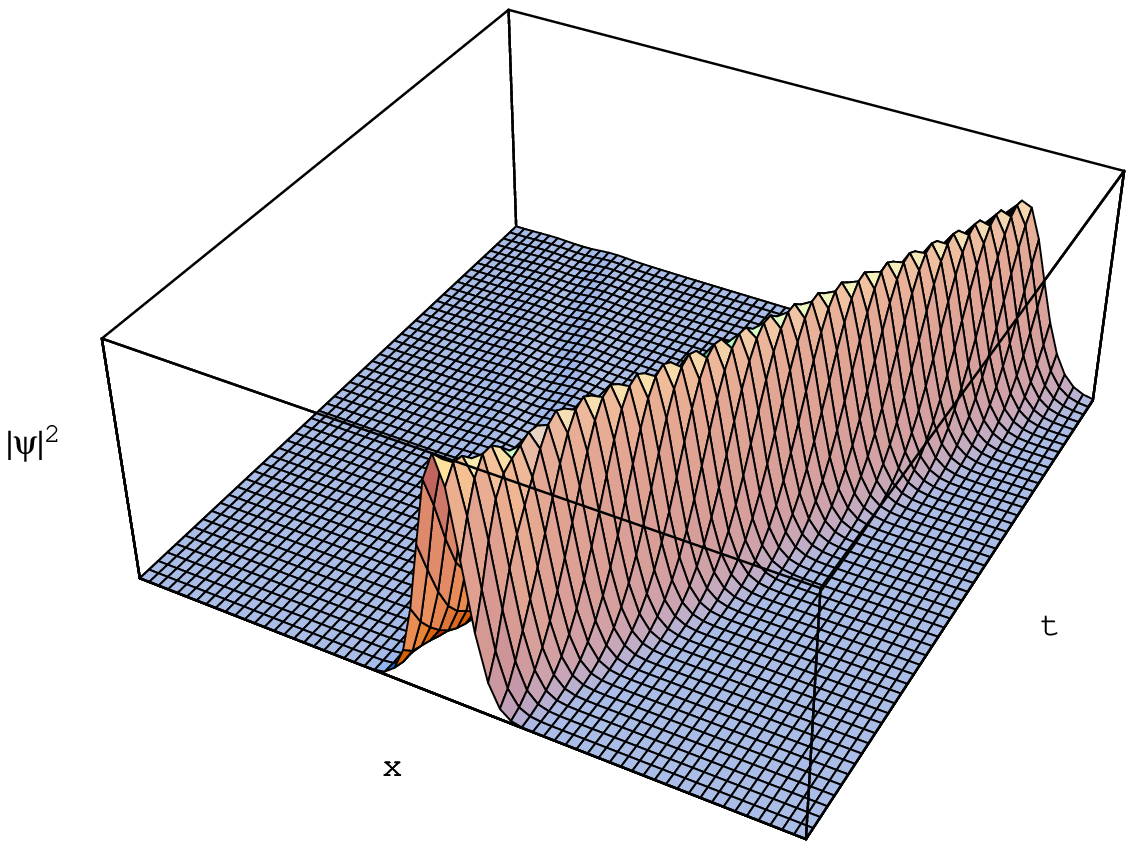}
$$
\vskip-\baselineskip
\eightpoint{%
{\narrower\noindent{\bf Figure 5}.  Evolution of the $k_0 = \pi/4$
wave packet (4.1) with width $s = 32$ for rule parameters 
$\theta = \pi/3$, $\rho = \pi/4$.  The probability peak moves from 
$x = 31$ at $t = 0$ to $x = 54$ at $t = 49$; thus the group 
velocity is approximately $23/49 \approx 0.47$.\par} 
}
\endinsert

This simulation shows that the $k_0 = \pi/4$ wave packet moves with 
well defined {\sl group velocity\/} to the right.  The result is just
what we would expect by analogy with the continuum situation and can
be analyzed in the same way, by transforming to the 
$|k,\epsilon\rangle$ basis.  Using (3.9) we compute the amplitudes in
this basis:
$$
\eqalignno{
\hat\psi_{\epsilon}(k) 
 &= \bigl(\psi^{(k,\epsilon)}(0)\bigr)^{\dagger}
    {2s \choose s}^{\!\! -1}\sum_x \psi^{(k_0,+1)}(0) e^{ik_0 x} 
                              {s \choose x - x_0 + s/2}e^{-ikx}    \cr
 &= \bigl(\psi^{(k,\epsilon)}(0)\bigr)^{\dagger}\psi^{(k_0,+1)}(0)
    {2s \choose s}^{\!\! -1}\sum_x {s \choose x - x_0 + s/2} 
                              e^{i(k_0 - k)x}                      \cr
 &= \bigl(\psi^{(k,\epsilon)}(0)\bigr)^{\dagger}\psi^{(k_0,+1)}(0)
    {2s \choose s}^{\!\! -1} e^{i(k_0 - k)(x_0 - s/2)} 
    (1 + e^{i(k_0 - k)})^s                                         \cr
 &= \bigl(\psi^{(k,\epsilon)}(0)\bigr)^{\dagger}\psi^{(k_0,+1)}(0)
    {2s \choose s}^{\!\! -1} e^{i(k_0 - k)x_0} 
    2^s \cos^s\bigl({k_0 - k \over 2}\bigr)                  &(4.2)\cr
}
$$
The amplitudes (4.2) give probabilities peaked around $k = k_0$, so 
this is also a wave packet in momentum space.  As usual, the group
velocity is the slope of the dispersion relation (3.5), \ie, 
${\rm d}\omega/{\rm d}k|_{k_0}$, which is 
$\sqrt{9 - 2\sqrt{6}}/4 \approx 0.49$ for the values used in the 
simulation of Figure 5; this is in good agreement with the measured
value of approximately 0.47.

The width of the peak in (4.2) depends inversely on $s$:  as $s$ 
decreases, \ie, the width of the wave packet in position space 
decreases, the width of the momentum peak increases.  The simulation
in Figure 6 shows the evolution of a wave packet with width $s = 8$.  
We note that while the group velocity is the same as in Figure 5, 
there is substantially more dispersion, indicating a greater interval 
of contributing wave numbers.  This is a general result, not depending 
on the specific form of our wave packet; the {\sl reciprocity 
relation\/} for the discrete Fourier transform has consequences 
similar to those of the uncertainty relation for the continuous 
Fourier transform [19].

\topinsert
\null\vskip-\baselineskip
\vskip-\baselineskip
\vskip-\baselineskip
$$
\epsfxsize=\usewidth\epsfbox{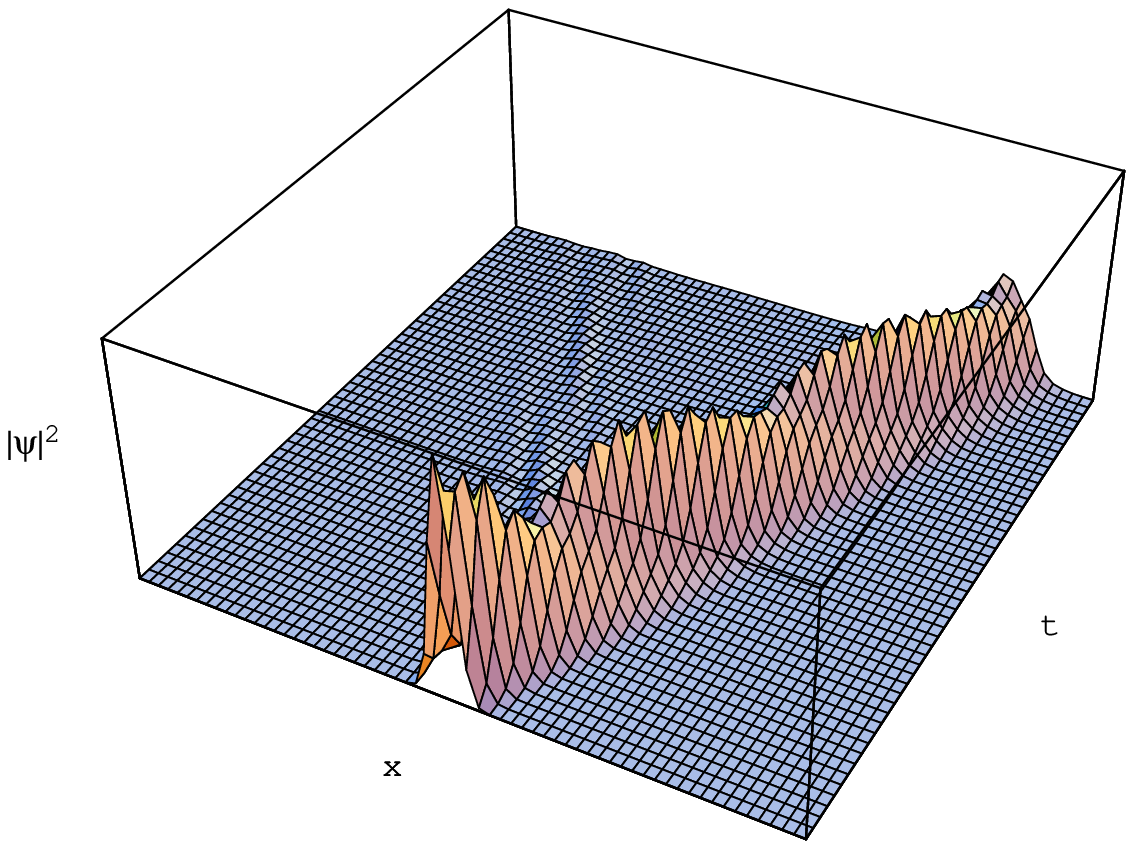}
$$
\vskip-\baselineskip
\eightpoint{%
{\narrower\noindent{\bf Figure 6}.  Evolution of the $k_0 = \pi/4$ 
wave packet (4.1) with width $s = 8$ for rule parameters 
$\theta = \pi/3$, $\rho = \pi/4$.  This wave packet disperses
more rapidly than the one shown in Figure 5:  The peak probability at
the end of the simulation is less than half of the initial peak
probability; left moving ripples carrying off some of the probability
are also visible.\par} 
}
\endinsert

Figure 7 shows a simulation of a wave packet built from the plane wave
with smallest nonzero wave number on the $\Z_{64}$ lattice:  
$k_0 = \pi/32$.  The horizontal tangent to the graph of the dispersion 
relation at $k = 0$, as shown in Figure 3, indicates that the group 
velocity of this wave packet will be small.  Furthermore, even with 
width $s = 32$ the wave number interval includes the left going modes
whose presence is visible in Figure 7; the consequence is an 
interference pattern and no very well defined group velocity.

\pageinsert
\null\vskip-\baselineskip
\vskip-\baselineskip
\vskip-\baselineskip
$$
\epsfxsize=\usewidth\epsfbox{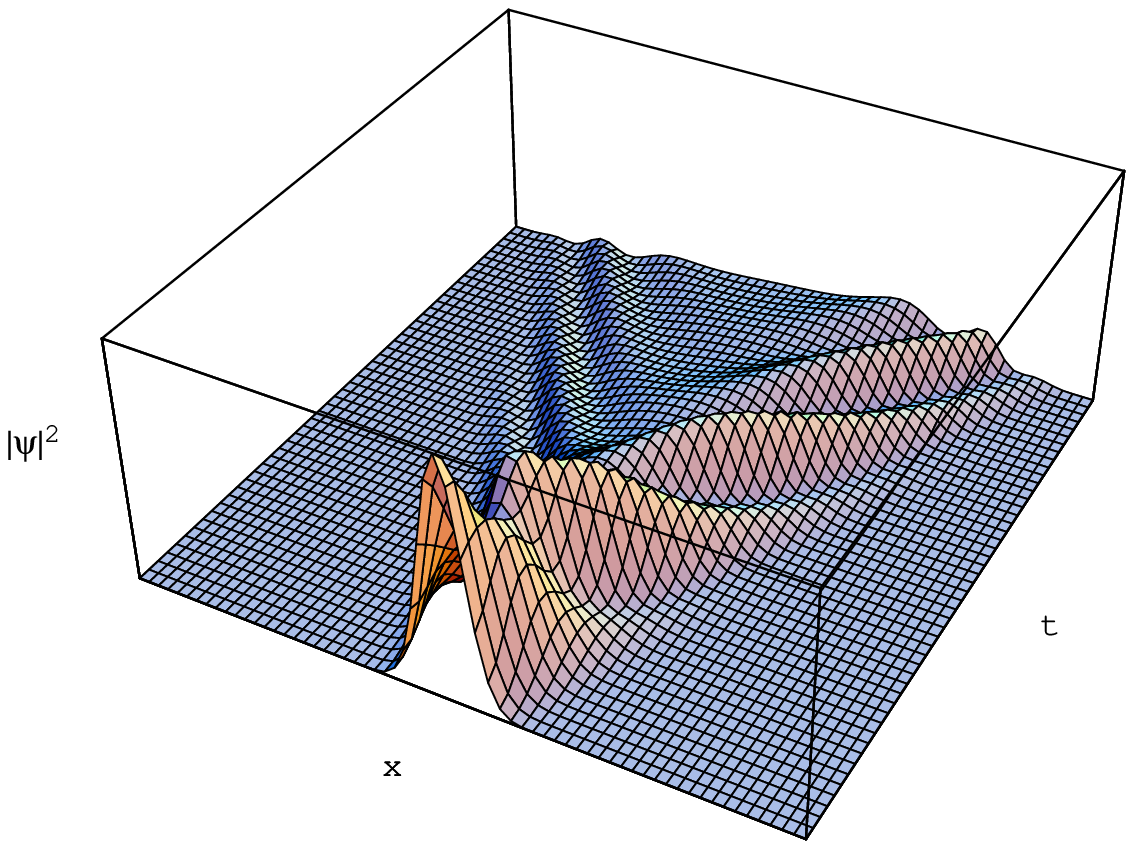}
$$
\vskip-\baselineskip
\eightpoint{%
{\narrower\noindent{\bf Figure 7}.  Evolution of the $k_0 = \pi/32$ 
wave packet (4.1) with width $s = 32$; the rule parameters are still
$\theta = \pi/3$, $\rho = \pi/4$.  This wave packet disperses
even more rapidly than the one shown in Figure 6:  left moving waves
carry off some of the probability and an interference pattern is
created.\par} 
}

\vfill
\null\vskip-\baselineskip
\vskip-\baselineskip
\vskip-\baselineskip
$$
\epsfxsize=\usewidth\epsfbox{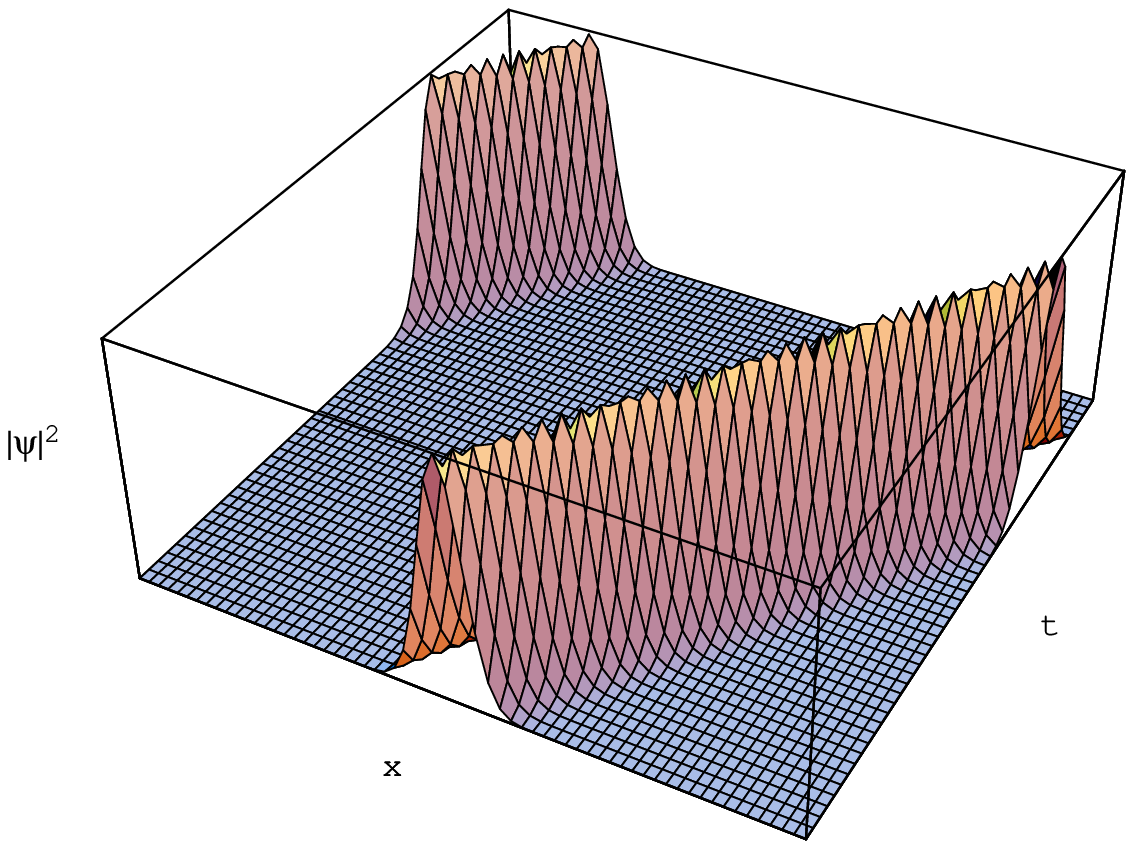}
$$
\vskip-\baselineskip
\eightpoint{%
{\narrower\noindent{\bf Figure 8}.  Evolution of the $k_0 = \pi/32$ 
wave packet (4.1) with width $s = 32$ for rule parameters 
$\theta = \pi/6 = \rho$.  This wave packet disperses very little and
has group velocity close to 1 in lattice units.\par} 
}
\endinsert

Finally, Figure 8 shows the evolution of the same wave packet but for
the rule parameters $\theta = \pi/6 = \rho$ whose dispersion relation
is graphed in Figure 4.  Here the group velocity is close to 1 in
lattice units, even for $k_0$ as small as $\pi/32$; the particle is 
indeed `massless'.  There is almost no dispersion in this simulation;
the probability contained in left going modes is nonzero, but too 
small by several orders of magnitude to be visible in Figure 8.

\medskip
\noindent{\bf 5.  Potentials}
\nobreak

\nobreak
\noindent The one particle QLGA described in Section 2 is the most
general {\sl homogeneous\/} model for particle speeds no more than 1.
To simulate physical systems (or to do useful computation), some
inhomogeneity must be introduced.  In each of the equations (2.6),
which express the unitarity condition, all the $w_i$ correspond to the 
scattering/interaction at a single lattice point, as do all the 
$w_i^{\dagger}$.  In the first equation these are the same lattice 
point, while in the second and third they are different.  Thus if 
$w_i(x) = w_i$, constants independent of $x$, solve these equations, 
so do $e^{-i\phi(x)}w_i$.  As observed already by Feynman [1] and 
Riazanov [2], such an $x$ dependent phase realizes an inhomogeneous 
potential in the continuum limit of the discrete path sum for the 
Dirac equation.  Here we investigate its effects on the quantum 
mechanics of our LGA, expecting them to be similar to those in the 
continuum limit.

For simplicity, we restrict our attention to a finite square well
potential, \ie, 
$$
w_i(x) := \cases{e^{-i\phi} w_i & if $N/4 \le x < 3N/4$;           \cr
                            w_i & otherwise,                       \cr
                }
$$
where the $w_i$ are defined by (2.7).  We begin by considering the 
effect of different values for $\phi$.

\moveright\secondstart\vtop to 0pt{\hsize=\halfwidth
\null\vskip-\baselineskip
\vskip-\baselineskip
\vskip-0.35\baselineskip
$$
\epsfxsize=\halfwidth\epsfbox{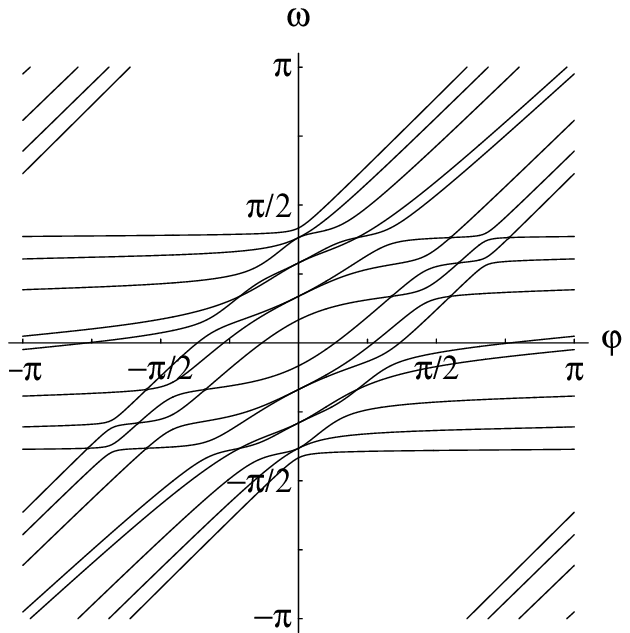}
$$
\eightpoint{%
\noindent{\bf Figure 9}.  The eigenvalues $\omega$ of $U$ for a square
well of depth $\phi$ and width $N/2$ on a lattice of size $N = 8$ with
$\theta = \pi/3$ and $\rho = \pi/4$.
}}
\vskip-\baselineskip
\parshape=16
0pt \halfwidth
0pt \halfwidth
0pt \halfwidth
0pt \halfwidth
0pt \halfwidth
0pt \halfwidth
0pt \halfwidth
0pt \halfwidth
0pt \halfwidth
0pt \halfwidth
0pt \halfwidth
0pt \halfwidth
0pt \halfwidth
0pt \halfwidth
0pt \halfwidth
0pt \hsize
Recall that the frequency (or energy) eigenvalues $\omega$ are doubly 
degenerate except for those with the largest and smallest absolute
value.  (See Figure 3, where each horizontal line intersecting the 
graph of the dispersion relation does so at two points except when
tangent to the maximum or minimum of either branch of the curve.)  As
with any perturbation to the evolution, we expect the introduction of
an inhomogeneity in the potential to resolve the degenerate 
eigenvalues.  Figure 9 shows that this is indeed the case:  as $\phi$
increases away from 0 the eigenvalues $\omega$ of $U$ increase and the
degenerate ones split.

The eigenvalues in Figure 9 have been computed for only $N = 8$; 
Figure 10 shows the results for $N = 32$.  On the larger lattice it is
clear what happens:  the horizontal bands of frequency/energy 
eigenvalues correspond to the eigenvalues of the unperturbed, 
homogeneous system, while the diagonal bands of eigenvalues correspond 
to the same ones, but shifted by the depth $\phi$ of the square well.  
The periodicity along the frequency axis shown in these graphs is a 
symptom of the ambiguity in the definition of energy due to discrete 
time evolution [20].  The graphs in Figures 9 and 10 have been 
computed for the QLGA with parameter values $\theta = \pi/3$, 
$\rho = \pi/4$, the dispersion relation for which is shown in Figure 3.  
Repeating the calculations for the `massless' case, with dispersion 
relation shown in Figure 4, gives the frequency/energy eigenvalue plot 
shown in Figure 11.  The degenerate levels still split, but much less 
than before for the same $\phi$ values, and the part of the band 
structure resulting from the nonzero minimum positive frequency in the 
massive dispersion relation vanishes.

\topinsert
\null\vskip-\baselineskip
\vskip-\baselineskip
$$
\epsfxsize=\halfwidth\epsfbox{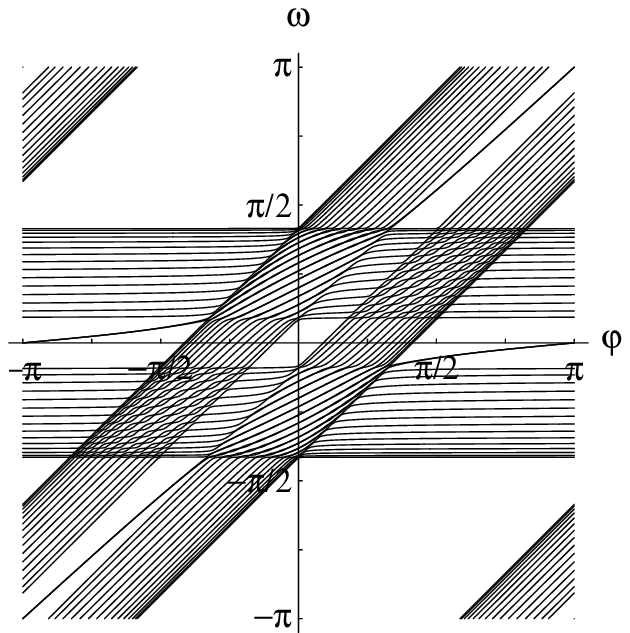}\hskip\chasm%
\epsfxsize=\halfwidth\epsfbox{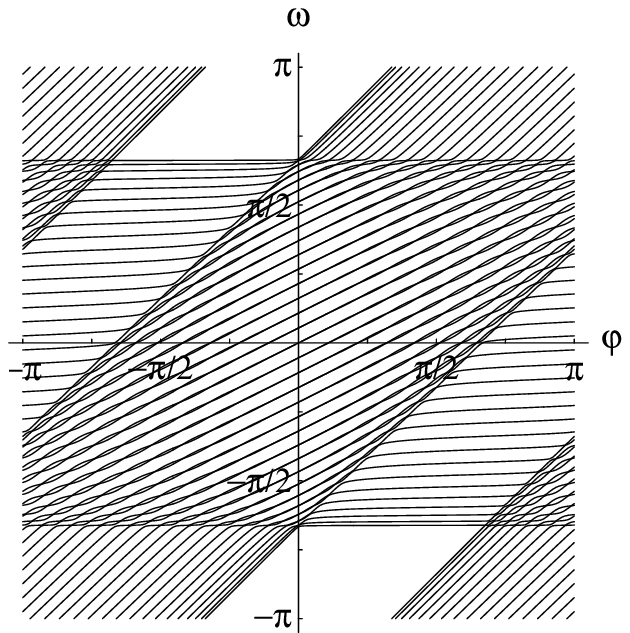}
$$
\hbox to\hsize{%
\vbox{\hsize=\halfwidth\eightpoint{%
\noindent{\bf Figure 10}.  The eigenvalues $\omega$ of $U$ for a 
square well of depth $\phi$ and width $N/2$ on a lattice of size 
$N = 32$ with $\theta = \pi/3$ and $\rho = \pi/4$.
}}
\hfill%
\vbox{\hsize=\halfwidth\eightpoint{%
\noindent{\bf Figure 11}.  The eigenvalues $\omega$ of $U$ for a 
square well of depth $\phi$ and width $N/2$ on a lattice of size 
$N = 32$ in the `massless' case  $\theta = \pi/6 = \rho$.
}}}
\endinsert

Now consider the eigenvectors of $U$, namely the eigenfunctions for 
our discrete version of a finite square well.  Since $U$ is no longer
translation invariant we do not have an equation like (3.3) to solve
for the eigenfunctions analytically.  Rather than developing a cross
boundary matching method as is used in the continuum problem for a 
periodic square well potential [21], here we simply find the 
eigenvectors of $U$ numerically.  Figure 12 shows the eigenfunctions 
corresponding to the three smallest positive eigenvalues for the QLGA 
with $\theta = \pi/3$ and $\rho = \pi/4$.  The depth of the square 
well is $\phi = \pi/24$.  We see exactly the lowest modes we would 
expect from our experience with such a potential in the continuum.  As 
the energy of the eigenfunction increases there is greater probability 
that the particle is outside the well---in the region of higher 
potential.  Figure 13 shows an eigenfunction which is approximately a 
plane wave in both regions:  it has larger wave number in the well 
than outside it.  For analytic results on the closely related problem 
of a step potential, and some discussion of their consequences for the
physical interpretation of QLGA, see [15].

\topinsert
\null\vskip-\baselineskip
\vskip-\baselineskip
$$
\epsfxsize=\halfwidth\epsfbox{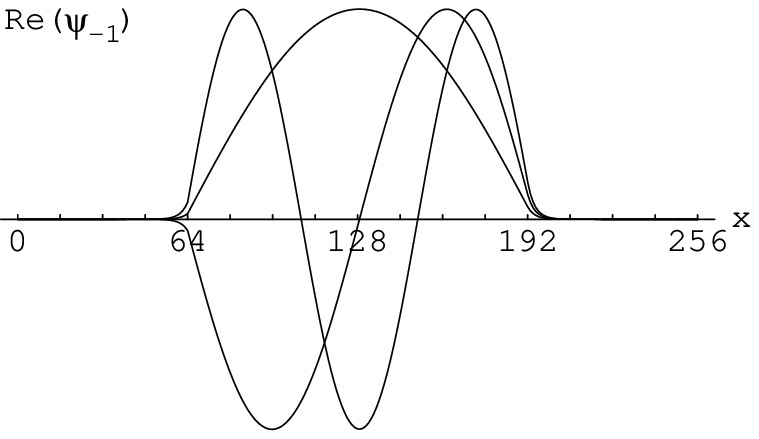}\hskip\chasm%
\epsfxsize=\halfwidth\epsfbox{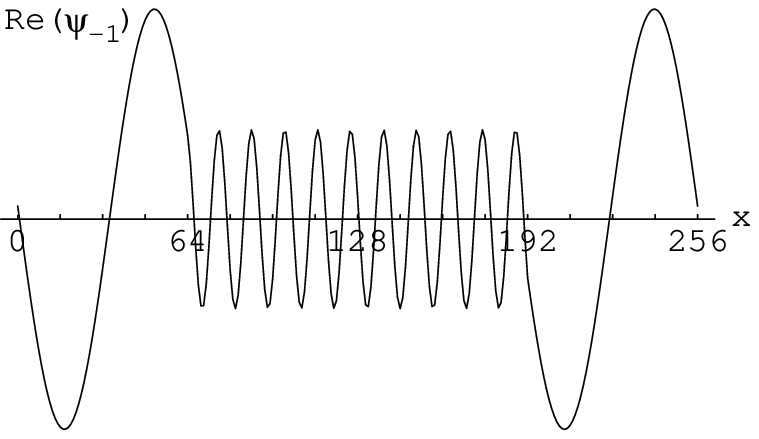}
$$
\hbox to\hsize{%
\vbox{\hsize=\halfwidth\eightpoint{%
\noindent{\bf Figure 12}.  The three eigenfunctions of $U$ with 
smallest positive eigenvalues:  $\pi/12 < 0.2622 < 0.2634 < 0.2653$,
for a square well of depth $\pi/24$ and width $N/2$ on a lattice of 
size $N = 256$ with $\theta = \pi/3$ and $\rho = \pi/4$.
}}
\hfill%
\vbox{\hsize=\halfwidth\eightpoint{%
\noindent{\bf Figure 13}.  An eigenfunction for a particle with
eigenvalue $0.3985 < 5\pi/12$ large enough not to be confined 
completely to the square well of the previous figure.  Outside the 
square well the wave number decreases and the probability increases.
}}}
\endinsert

Finally, suppose we prepare one of the semiclassical wave packets 
studied in Section 4 in a finite square well.  Using the dispersion 
relation (3.5), we find that the $k_0 = \pi/4$, width $s = 32$ wave 
packet (4.1) of Figure 5 has peak frequency 
$\omega_0 = {\rm Cos}^{-1}\bigl((1 + \sqrt{6})/4\bigr)$ for rule 
parameters $\theta = \pi/3$, $\rho = \pi/4$.  
$\omega_0$ is just a little larger than $\pi/6$ so we would not expect
a square well of depth $\phi = \pi/6$ to contain this wave packet.  
Figure 14 shows a simulation of this situation on a lattice of size 
$N = 64$:  the wave packet continues past the right edge of the square 
well at $3N/4$ with only a small amount of internal reflection.

\midinsert
\null\vskip-9\baselineskip
$$
\epsfxsize=\usewidth\epsfbox{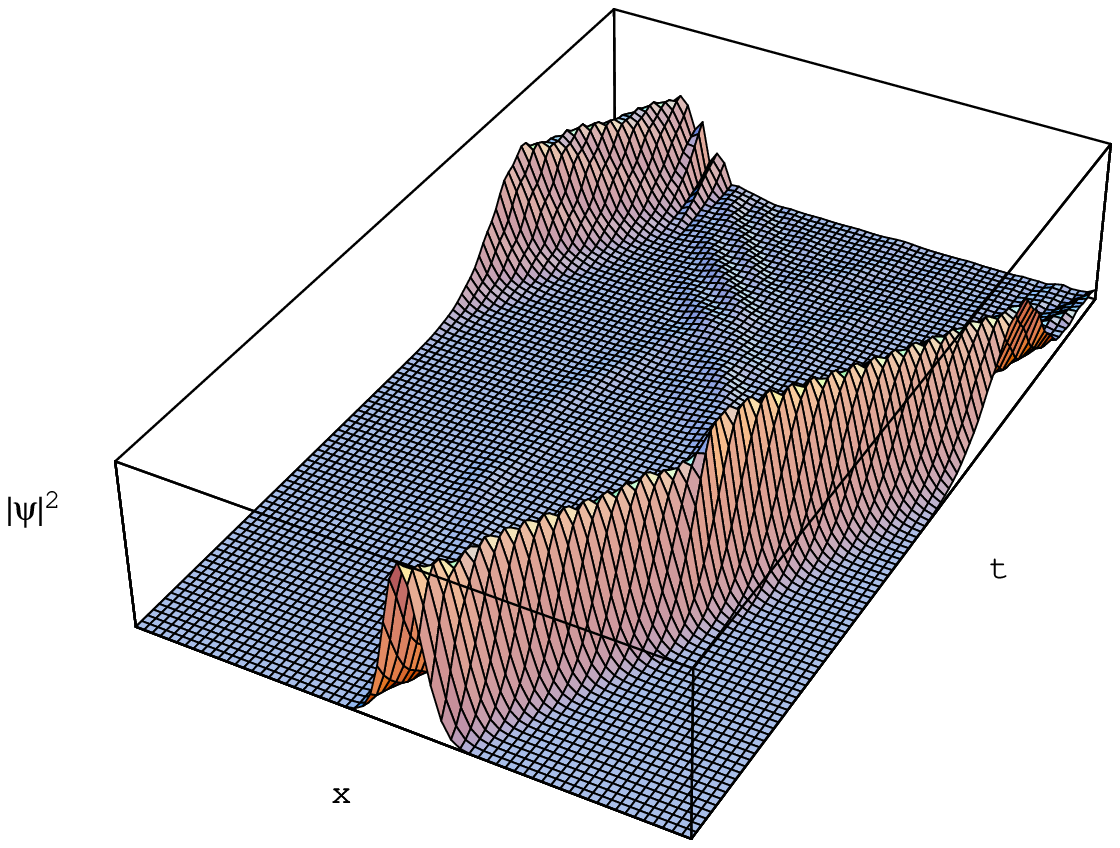}
$$
\vskip-7\baselineskip
\eightpoint{%
{\narrower\noindent{\bf Figure 14}.  Evolution of the $k_0 = \pi/4$ 
wave packet (4.1) with width $s = 32$ for rule parameters 
$\theta = \pi/3$, $\rho = \pi/4$ in a square well of depth 
$\phi = \pi/6$.  There is very little reflection as the wave packet
passes the right wall of the square well at $x = 3N/4 = 48$.\par} 
}
\endinsert

\pageinsert
\null\vskip-9\baselineskip
\vfill
\centerline{({\it figure available from author\/})}
\vfill
\vskip-6\baselineskip
{\eightpoint{%
{\narrower\noindent{\bf Figure 15}.  Evolution of the same wave packet 
with the same rule parameters as in Figure 14, but now in a square 
well of depth $\phi = \pi/4$.  There is both reflection and 
transmission as the wave packet scatters off the right wall of the 
square well.\par} 
}}

\vfill
\null\vskip-9\baselineskip
\centerline{({\it figure available from author\/})}
\vfill
\vskip-8\baselineskip
\eightpoint{%
{\narrower\noindent{\bf Figure 16}.  Evolution of the same wave packet 
with the same rule parameters as in Figures 14 and 15, but now in a 
square well of depth $\phi = \pi/3$.  This well is deep enough that 
the wave packet is almost entirely reflected by the right wall of the 
square well.\par} 
}
\endinsert

Increasing the depth of the square well should have the effect of
increasing the amount of internal reflection of the wave packet.  
Simulations demonstrate that this is indeed the case.  When the depth 
of the square well is $\phi = \pi/4$, Figure 15 shows that the wave
packet splits as it scatters off the right wall of the square well.
With greater probability the particle is reflected back into the well,
but it also has a substantial probability of continuing to the right.
The wave packet which continues to the right does so at a reduced 
group velocity as we can see by the fact that the reflected wave 
packet travels back across the well, reaching the left wall at 
$x = N/4 = 16$ at the end of the simulation shown, before the 
transmitted wave packet travels the same distance rightwards.

Finally, when the depth of the square well is increased to 
$\phi = \pi/3$, Figure 16 shows that almost the entire wave packet is
reflected back into the well by the right wall.  In this case there is
only a very small probability that the particle has sufficiently large
energy to escape the well.

\medskip
\noindent{\bf 6.  Discussion}
\nobreak

\nobreak
\noindent Unitarity is a very restrictive constraint on the local 
scattering rule for a QLGA with a single particle of bounded speed.
When the bound is 1 in lattice units, there is a two parameter family
of reflection invariant one dimensional local rules, given by (2.5)
and (2.7).  It is already remarkable that the Dirac equation arises as
a continuum limit of this QLGA when $\rho = 0$.  In this paper we have
begun to investigate the quantum mechanics of the general two 
parameter rule.  We find that {\sl even without going to a continuum 
limit\/} the QLGA reproduces the quantum mechanical phenomena of plane 
waves and wave packets obeying a dispersion relation (3.5).  
Furthermore, the model straightforwardly accommodates the inclusion of 
inhomogeneous potentials.  The eigenvectors of the evolution matrix 
give the quantum mechanical eigenfunctions for the lattice gas 
particle, and simulations exhibit the semi-classical evolution of a 
wave packet, in the presence of a square well potential.

Taking a QLGA seriously as a possible model for quantum computation
by, for example, ballistic electrons in a lattice of solid state
nanostructures, raises many additional questions, some of which will 
be addressed in subsequent papers in this series:  Inhomogeneity of 
the substrate can be incorporated in the model by varying the rule
parameters while maintaining global unitarity.  Finite, non-periodic,
boundary conditions can be imposed similarly [22].  Higher 
dimensional [2,4,10,14] and multiparticle [6,14,17] models can also be 
constructed.  Decoherence is the crucial problem for quantum computers 
[8], particularly in the solid state [23].  QLGA provide an extremely 
convenient arena in which to model this problem [24].  Finally, the 
question of for which quantum computational tasks QLGA are best suited 
deserves serious investigation.

\vfill\eject

\noindent{\bf Acknowledgements}
\nobreak

\nobreak
\noindent It is a pleasure to thank Peter Doyle, Mike Freedman, 
Peter Monta and Richard Stong for discussions on various aspects of 
this project.  I also benefited from conversations with Herbert 
Bernstein, Thomas Beth, P\"aivi T\"orm\"a and Harald Weinfurter at the
$4^{\rm\eightpoint th}$ Workshop on Quantum Computation sponsored by
the ISI Foundation.

\global\setbox1=\hbox{[00]\enspace}
\parindent=\wd1

\noindent{\bf References}
\bigskip

\parskip=0pt
\item{[1]}
\feynman\ and A. R. Hibbs,
{\sl Quantum Mechanics and Path Integrals}
(New York:  McGraw-Hill 1965).

\item{[2]}
G. V. Riazanov,
``The Feynman path integral for the Dirac equation'',
\SPJETP\ {\bf 6} (1958) 1107--1113.

\item{[3]}
S. Succi and R. Benzi,
``Lattice Boltzmann equation for quantum mechanics'',
\PD\ {\bf 69} (1993) 327--332;\hfb
S. Succi,
``Numerical solution of the the Schr\"odinger equation using 
  discrete kinetic theory'',
\PRE\ {\bf 53} (1996) 1969--1975.

\item{[4]}
I. Bialynicki-Birula,
``Weyl, Dirac, and Maxwell equations on a lattice as unitary 
  cellular automata'',
\PRD\ {\bf 49} (1994) 6920--6927.

\item{[5]}
\gz,
``Quantum cellular automata'',
\CS\ {\bf 2}\break
(1988) 197--208;\hfb
S. Fussy, G. Gr\"ossing, H. Schwabl and A. Scrinzi,
``Nonlocal computation in quantum cellular automata'',
\PRA\ {\bf 48} (1993) 3470--3477;\hfb
and references therein.

\item{[6]}
\dajm,
``From quantum cellular automata to quantum lattice gases'',
UCSD preprint (1995), quant-ph/9604003, to appear in \JSP

\item{[7]}
W. D. Hillis,
``New computer architectures and their relationship to physics or
  why computer science is no good'',
\IJTP\ {\bf 21} (1982) 255--262;\hfb
N. Margolus,
``Parallel quantum computation'',
in W. H. Zurek, ed.,
{\sl Complexity, Entropy, and the Physics of Information},
proceedings of the SFI Workshop, Santa Fe, NM, 
29 May--10 June 1989,
{\sl SFI Studies in the Sciences of Complexity} {\bf VIII}
(Redwood City, CA:  Addison-Wesley 1990) 273--287;\hfb
\brosl,
``Parallel billiards and monster systems'',
in N. Metropolis and G.-C. Rota, eds.,
{\sl A New Era in Computation}
(Cambridge:  MIT Press 1993) 53--65;\hfb
R. Mainieri,
``Design constraints for nanometer scale quantum computers'',
preprint (1993) LA-UR 93-4333, cond-mat/9410109;\hfb
M. Biafore,
``Cellular automata for nanometer-scale computation'',
\PD\ {\bf 70}\break
(1994) 415--433.

\item{[8]}
D. P. DiVincenzo,
``Quantum computation'',
\Sc\ {\bf 270} (1995) 255--261;\hfb
I. L. Chuang, R. Laflamme, P. W. Shor and W. H. Zurek,
``Quantum computers, factoring, and decoherence'',
\Sc\ {\bf 270} (1995) 1633--1635;\hfb
A. Barenco and A. Ekert,
``Quantum computation'',
\APS\ {\bf 45} (1995) 1--12;\hfb
and references therein.

\vfill\eject

\item{[9]}
\feynman,
``Simulating physics with computers'',
\IJTP\ {\bf 21} (1982) 467--488.

\item{[10]}
\dajm,
``On the absence of homogeneous scalar unitary cellular automata'',
UCSD preprint (1995), quant-ph/9604011, to appear in \PLA.

\item{[11]}
C. D\"urr, H. L\^e Thanh and M. Santha,
``A decision procedure for well-formed linear quantum cellular 
  automata'',
in C. Puecha and R. Reischuk, eds.,
{\sl STACS 96:  Proceedings of the 13th Annual Symposium on 
     Theoretical Aspects of Computer Science},
Grenoble, France, 22--24 February 1996,
{\sl Lecture notes in computer science} {\bf 1046}
(New York:  Springer-Verlag 1996) 281--292;\hfb
C. D\"urr and M. Santha,
``A decision procedure for unitary linear quantum cellular
automata'',
preprint (1996), quant-ph/9604007;\hfb
\dajm,
``Unitarity in one dimensional nonlinear quantum cellular automata'',
UCSD preprint (1996), quant-ph/9605023;\hfb
W. van Dam,
``A universal quantum cellular automaton'',
preprint (1996).

\item{[12]}
\feynman,
``Quantum mechanical computers'',
\FP\ {\bf 16} (1986) 507--531.

\item{[13]}
P. Benioff,
``Quantum ballistic evolution in quantum mechanics:  application
  to quantum computers'',
preprint (1996), quant-ph/9605022.

\item{[14]}
B. M. Boghosian and W. Taylor, IV,
``A quantum lattice-gas model for the many-particle Schr\"odinger
  equation in $d$ dimensions'',
preprint (1996) BU-CCS-960401, PUPT-1615, quant-ph/9604035.

\item{[15]}
\dajm,
``Quantum lattice gases and their invariants'',
UCSD preprint.

\item{[16]}
P. A. M. Dirac,
{\sl The Principles of Quantum Mechanics}, fourth edition
(Oxford:  Oxford University Press 1958).

\item{[17]}
H. Weyl,
{\sl The Theory of Groups and Quantum Mechanics},
translated from the second (revised) German edition by H. P. 
 Robertson
(New York:  Dover 1950).

\item{[18]}
G. Zanetti,
``Hydrodynamics of lattice gas automata'',
\PRA\ {\bf 40} (1989) 1539--1548;\hfb
Z. Cheng, J. L. Lebowitz and E. R. Speer,
``Microscopic shock structure in model particle systems:  The
  Boghosian-Levermore cellular automaton revisited'',
\CPAM\ {\bf XLIV} (1991) 971--979;\hfb
\bd,
``Lattice gases and exactly solvable models'',
\JSP\ {\bf 68} (1992) 575--590.

\item{[19]}
W. L. Briggs and V. E. Henson,
{\sl The DFT:  An owner's manual for the discrete Fourier transform\/}
(Philadelphia:  SIAM 1995).

\item{[20]}
G. 't Hooft,
``Equivalence relations between deterministic and quantum mechanical
  systems'',
\JSP\ {\bf 53} (1988) 323--344.

\item{[21]}
J. C. Slater,
{\sl Quantum Theory of Matter\/}
(New York:  McGraw-Hill 1951) Chapter 10.

\item{[22]}
\dajm,
``Quantum mechanics of lattice gas automata.  II.  Boundary 
  conditions and other inhomogeneities'',
UCSD/IPS preprint.

\vfill\eject

\item{[23]}
T. Ando, A. B. Fowler and F. Stem,
``Electronic properties of two-dimensional systems'',
\RMP\ {\bf 54} (1982) 437--672;\hfb
D. K. Ferry and H. L. Grubin,
``Modeling of quantum transport in semiconductor devices'',
in H. Ehrenreich and F. Spaepen, eds.,
{\sl Solid State Physics:  Advances in Research and Applications\/}
{\bf 49} (1995) 283--448;\hfb
and references therein.

\item{[24]}
\dajm,
in preparation.

\bye